\documentclass[aps,prc,twocolumn,groupedaddress,showpacs]{revtex4-1}
\usepackage{color,soul}
\usepackage{graphicx}
\usepackage{amsmath}
\usepackage{cancel}
\usepackage{xcolor}

\newcommand {\la} {\langle}\newcommand {\ra} {\rangle}
\newcommand {\beq} {\begin{eqnarray}}
\newcommand {\eeq} {\end{eqnarray}}
\newcommand {\eeqn} [1] {\label{#1} \end{eqnarray}}
\newcommand {\eol} {\nonumber \\}
\newcommand {\ve} [1] {\mbox{\boldmath $#1$}}
\newcommand {\dem} {\mbox{$\frac{1}{2}$}}

\begin{document}

\title{Magnetic dipole contribution to the $\gamma$-spectrum from $d$-$t$  collisions}

\author{N. K. Timofeyuk}
\affiliation{School  of Mathematics and Physics, Faculty of Engineering and Physical
Sciences, University of Surrey, Guildford, Surrey, GU2 7XH, United Kingdom}

\date{\today}

\begin{abstract}

One in about a few hundred thousand sub-Coulomb $d$-$t$ collisions is accompanied by the emission of a $\gamma$-photon. The $\gamma$-spectrum from these collisions is  dominated by a  16.7 MeV peak, corresponding to the population of the $p$-wave $3/2^-$ $\alpha$-$n$ ground-state resonance   in  the E1 transition from an intermediate $d$-wave $\alpha$-$n$ state, corresponding to the $^5$He excited state around 16.7 MeV. The strength of this spectrum at the  large $\gamma$-energy endpoint decreases fast both due to kinematic factors and due to centrifugal  repulsion between neutron and $\alpha$-particle at near-zero relative energies. However, no centrifugal repulsion would occur if $s$-wave $\alpha$-$n$ states were populated in magnetic dipole transition. Therefore, one can expect that close to the $\gamma$-spectrum endpoint around 17.5 MeV the M1 transition could dominate, leading to additional counts in experimental spectra which could be easily misinterpreted as a background.
This work presents  calculations of the M1 contribution to the $d+t\rightarrow \alpha+n+\gamma$  cross section caused both by convection and magnetization currents. While the first contribution was found to be negligible, the one from magnetization depends strongly on the choice of the model for the $d$-$t$ channel scattering wave function. Using  $d$-$t$ interaction models from the literature, represented either by a gaussian of by a square well, resulted in   unrealistically strong M1 contribution. With  constructed in the present work  optical folding potential, correctly representing the sizes of deuteron and triton, the M1 contribution near the endpoint is reduced to about 2$\%$ of the 16.7-MeV E1 peak. Arguments in favor of more detailed calculations, that could potentially change this number, are put forward. 
\end{abstract}

\maketitle

\section{Introduction}

The $d+t\rightarrow \alpha+n$ reaction is the best candidate for successful fusion-energy-exploiting industrial applications. Rarely, for about one in a few hundred thousand events, the $d$-$t$ collisions end up with the emission of a $\gamma$-ray. The $\gamma$-spectrum has a peak at around 16.7 MeV, corresponding to the population of the $3/2^-$ ground-state resonance in the $\alpha$-$n$ system.
Given that these high-energy $\gamma$'s could potentially become a basis of a new $d$-$t$ plasma diagnostic tool \cite{Mor92}, the study of the $d+t\rightarrow \alpha+n+\gamma$ reaction has attracted significant interest from the experimental nuclear physics community. Experimental studies of this reaction were carried out in cyclotrons through 1960-1990s \cite{Bus63,Kos70,Bez70,Cec84,Mor86,Kam93} and since 2010s they are also conducted in inertial confinement fusion (ICF) facilities \cite{Kim12, Jee21,Mea21,Moh22,Moh23,Mea24}. The first results on $\gamma$-detection in JET has just been reported in \cite{Dal24,Reb24}.

Until very recently, no theoretical predictions were available for the $d+t\rightarrow \alpha+n + \gamma$ to $d+t\rightarrow \alpha+n$  branching  ratio and the only available modelling tool was the
phenomenological R-matrix approach, which was used in \cite{Mea21, Hor21}
for $\gamma$-spectrum predictions  on the assumption that spectral features in the $\alpha$-$n$ scattering and the
$d+t\rightarrow \alpha+n+\gamma$ reaction  are exactly the same. 
The first theoretical prediction 
of the low-energy $d+t\rightarrow \alpha+n+\gamma$ reaction has been reported in \cite{our},   aimed at understanding both the shape of the $\gamma$-spectrum   and its absolute strength. The main challenge encountered in that work was   that the   widely-used in low-energy nuclear physics approximation for electric dipole transition amplitude, ${\cal M}^{E1} \approx \sum_i e_i \ve{r}_i$, forbids the  $d+t\rightarrow \alpha+n+\gamma$ reaction from happening  because of the spin conservation  rule. In the vicinity of the   $3/2^+$ resonance, determining the main properties  of the $d$-$t$ system, the deuteron spin $J_d  =1$ and the triton spin $J_t = 1/2$ couple to the total spin $\vec{S} = \vec{J}_d + \vec{J}_t$ equal to 3/2  while the channel spin in the $\alpha$-$n$ channel is different, being 1/2. However, the $d$-$t$ component of the $^5$He five-body wave function is coupled by the nucleon-nucleon tensor interaction to the high-energy $d$-wave $\alpha$-$n$ component  with the spin $S=1/2$ and it is the E1 $\gamma$-transition between the high- and low-energy $S=1/2$ states in the $\alpha$-$n$ partition of $^5$He that provides the main contribution to the $d+t\rightarrow \alpha+n+\gamma$ reaction mechanism.

If allowed by selection rules, electric dipole transitions are the strongest, with other multipolarities being suppressed. In the case of the  $d+t\rightarrow \alpha+n+\gamma$ reaction,
the electric quadrupole transitions E2  are  several orders of magnitude weaker than the E1 transitions \cite{our}. Usually, magnetic dipole transitions M1 have a similar    strength to that of the E2 ones. However, for the  $d+t\rightarrow \alpha+n+\gamma$ reaction these transitions should be studied more carefully because of the following reason. Since the M1 transition does not conserve the total spin of the system, it can directly populate the final $\alpha$-$n$ state from the   $d$-$t$ partition of $^5$He %that contains the 3/2$^+$ resonance, 
without any need for going through the high-energy $d$-wave  $\alpha$-$n$ state with $S=1/2$. The populated by M1 final $\alpha$-$n$ state has zero orbital momentum and, therefore, the same (positive) parity that the initial state has and thus will not contribute to the (negative parity) resonant peak featuring in the E1 $dt\gamma$-spectrum. However, it may contribute  to the high-energy endpoint of this spectrum because, unlike in the case of the $p$-wave $\alpha$-$n$ scattering, the $s$-wave motion is not affected by the centrifugal barrier and  the $\alpha$-$n$ wave function does not vanish at small $\alpha$-$n$ separations at  near-zero  energies. Also, the strength of the M1 transitions grows with $\gamma$-energy. As the result, some M1-originated counts could appear in the endpoint of the $d+t\rightarrow \alpha+n+\gamma$ spectrum, which could be easily misidentified as background.  Subtracting their contribution could induce a systematic error into the $\gamma$-spectrum presented as a measured quantity. 

In this paper   the M1 contribution to the $d+t\rightarrow \alpha+n+\gamma$ reaction is calculated using a potential model description of the $d$-$t$ and $\alpha$-$n$ channel functions. These calculations are based on the exact M1 operator, which along with $s$-wave to $s$-wave transitions, caused by magnetization currents, allows convection-induced bremsstrahlung  between $d$-wave states of the $^5$He. Sect. II  presents the M1 formalism for the case of the $d+t\rightarrow \alpha+n+\gamma$ reaction. Sect. III shows numerical results while Sect. IV discusses the findings made in the previous section and draws the conclusions.

\section{Formalism for $M1$-induced $d+t\rightarrow \alpha+n+\gamma$ transition}

The general form of the $M\lambda$ operator describing the interaction of a photon with an $A$-body system in first-order perturbation  theory is given  by the sum of two terms \cite{Bay12}:
\beq
{\cal M}^{M\lambda}_{\mu} &=& {\cal M}^{M\lambda; \, \rm conv}_{\mu} + {\cal M}^{M\lambda; \, \rm magn}_{\mu}
\eol
&=&
\frac{e\hbar}{m_p c} \frac{(2\lambda+1)!!}{(\lambda+1) k_{\gamma}^{\lambda}} \,\,
\sum_{j=1}^{A} 
\left\{ g_{l_j} \left[ \vec{\nabla}_j \phi_{\lambda \mu}(\ve{r}_j)\right] \cdot \vec{L}_j     \right.
\eol
 & & + \left. 
\frac{1}{2} g_{s_j}
\frac{\vec{\nabla}_j \times \vec{L}_j}{i} \phi_{\lambda \mu}(\ve{r}_j) \cdot 
\vec{S}_j, 
\right\},\,\,\,\,\,\,\,\,\,\,\,
\eeqn{operatorM}
where $e$ is the proton charge, $m_p$ is the proton mass, $g_{l_j}= \left(\frac{1}{2}-t_{3j}\right)$ with $t_{3j}$ being the isospin projection  of nucleon $j$ onto the $z$-axis, 
$g_{s_j} = (g_n+g_p)/2+t_{3j}(g_n-g_p)$ and $g_n$ and $g_p$ are gyromagnetic factors of the neutron and proton, respectively. Also, $\vec{L}_j$ is the operator of the orbital momentum of nucleon $j$,
\beq
\vec{L}_j = -i \vec{r}_j \times \vec{\nabla}_j,
\eeqn{Lj}
$\vec{S}_j$ is the spin operator acting on nucleon $j$ and 
\beq
\phi_{\lambda\mu}(\ve{r}_j)= j_\lambda(k_{\gamma}r_j)Y_{\lambda\mu}(\hat{r}_j) .
%= \frac{i^{-\lambda}}{4\pi}   \int d\hat{k}_{\gamma}  \, Y_{\lambda \mu}(\hat{k}_{\gamma}) e^{i \vec{k}_{\gamma}\vec{r}_j}, \,\,
\eeqn{philm}
%where $\phi_j = e^{i \vec{k}_{\gamma}\vec{r}_j}$. 
The coordinate  $\ve{r}_j$ of nucleon $j$ is with respect to the centre of mass of system $A$, which is important for such light systems as   $d+t$.

%The integral representation over angular part of $\vec{k}_{\gamma}$ in (\ref{philm}) is very useful for using various coordinate transformations to facilitate evaluation of (\ref{ME}). 

%The M1 operator contain $\sum_{i=1}^{5} \vec{S}_i$.  We first evaluate the direct term. To do this, we separate triton or $^3$He for $\alpha$ and deal with different nucleon separately.

The first term  of this operator, ${\cal M}^{M\lambda; \, \rm conv}_{\mu}$, comes from the convection current, arising from the motion of protons, while the second term, ${\cal M}^{M\lambda; \, \rm magn}_{\mu}$, is due to magnetization current, depending on the spins of all nucleons. The first term conserves the total nuclear spin, and therefore will contribute to bremsstrahlung between the $d$-wave states in the $\alpha$-$n$ partition of $^5$He. The second term can couple the $s$-wave $d$-$t$ and $\alpha$-$n$ partition. The formal expressions for differential cross section corresponding to convection and magnetization currents are considered below separately. The cross sections are calculated from the modulus squared of the matrix element $ \la \Psi_f | {\cal M}^{M1}_{\mu} |\Psi_i\ra $ between the wave functions $\Psi_i$ and $\Psi_f$ of the initial and final states of the $^5$He system.

\subsection{Partial wave representation of the wave functions}

The wave function $\Psi_f$ in the final channel is taken as an antisymmetrized product of the  4+1 partition wave function $\Phi^{(41)}(1,2,3,4;5)$ in which nucleons 1, 2, 3 and 4 belong to the $\alpha$-particle and nucleon $5$ is a neutron:
\beq
\Psi_f (1,2,3,4,5) 
&=& %{\cal A}_{dt} \Phi^{(32)}(1,2,3;4,5)
%\eol
%&+& 
{\cal A}_{\alpha n} \Phi^{(41)}_{n \alpha, M_n}(1,2,3,4;5),
\eeqn{Psi12345}
where ${\cal A}_{c_1 c_2}$ is the antisymmetrization operator
\beq 
 {\cal A}_{c_1 c_2} =\left(\frac{c_1! c_2! }{(c_1+c_2)!}\right)^{1/2} \sum_{\Pi_{c_1c_2}} \varepsilon_{\Pi_d} \Pi_{c_1c_2}, 
\eeqn{Anti}
 that contains all permutations ${\Pi_{c_1c_2}}$  between nucleons from two different clusters $c_1$ and $c_2$ and the sign  $\varepsilon_{\Pi_{c_1c_2} }$ accompanying these permutations, being either $+$ or $-$ depending on whether the number of permutations is even or odd.
%The channel wave functions $\Phi^{(32)}(1,2,3;4,5)$ and $\Phi^{(41)}(1,2,3,4;5)$ are  antisymmetric with respect to permutation of nucleons belonging to partitions separated by semicolons. 
The partition function  $\Phi^{(41)}_{n\alpha,M_n}(1,2,3,4;5)$ depends on the $\alpha$-$n$ momentum $ k_f$ and on the spin projection  $M_n$ of the neutron when it is far from $\alpha$-particle. It is represented by a general partial wave expansion of the  form 
%using Eqs. (4.66a,b) from  \cite{satchler} we have
\beq
\Phi^{(41)}_{n\alpha, M_n}(1,2,3,4;5) = \frac{1}{\sqrt{v_{f}}}\frac{4\pi}{k_{f}r} \, \psi_{\alpha}(1,2,3,4) \,\,\,\,\,\,\,\,\,\,\,\,\,\,\,
\,\,\,\,\,\,\,\,\,\,\,\,\,\,\,
\eol
\times 
%\sum_{\sigma_n}  
\sum_{
{\scriptsize \begin{array}{ll} 
 {L'_fL_fJ_f} \\ M_fM_{J_f}\sigma_{n_f}
    \end{array}  } }
%\begin{L'_fL_fJ_fM_fM_{J_f}\sigma_n}
(L_f M_f \dem M_n | J_f M_{J_f}) (L'_f M'_f \dem \sigma_{n_f} | J_f M_{J_f})
\eol
 \times 
 \chi^{(\rm spin)}_{\dem \sigma_n}  (5) \, i^{L'_f} Y^*_{L_fM_f}(\hat{\ve{k}}_{f}) Y_{L'_fM'_f}(\hat{\ve{r}}_{\alpha n})  \chi^{J_f}_{L_fL'_f} (k_{f},r_{\alpha n}), \,\,\,\,\,\,\,\,\,\,\,\,\, 
\eeqn{chif}
where  $v_f = \hbar k_f/\mu_{\alpha n}$ is the velocity in the exit $\alpha$-$n$ channel determined by the reduced mass $\mu_{\alpha n}$ and the energy $E_f = \hbar^2 k_f^2/  2\mu_{\alpha n} $, the $L_f$ and $J_f$ are orbital momentum and total angular momentum, respectively, and $M_f$ and $M_{J_f}$ are their corresponding projections. The Clebsch-Gordan coefficients provide angular momentum conservation, $\vec{L}_f + \vec{\frac{1}{2}} = \vec{J}_f$.
Also in (\ref{chif}), $\psi_{\alpha}(1,2,3,4)$ is the intrinsic antisymmetric $\alpha$-particle wave function, $\chi^{(\rm spin)}_{\dem \sigma_{n_f} } (5)$ is the spin wave function of the neutron{, labelled by number 5, with the spin projection $ \sigma_{n_f} $, and $ \chi^{J_f}_{L_fL'_f} (k_{f},r_{\alpha n})$ is the radial wave function that depends on the neutron coordinate with respect to the centre of mass of the $\alpha$-particle, $\ve{r}_{\alpha n} = \ve{r}_n - \ve{r}_{\alpha}$.

The wave function $\Psi_i$ in the entrance $d$-$t$ channel is represented by two partitions,
\beq
\Psi_i (1,2,3,4,5) 
&=& {\cal A}_{dt} \Phi^{(32)}_{dt, M_d M_t}(1,2,3;4,5)
\eol
&+& {\cal A}_{n\alpha, M_n} \Phi^{(41)}_{n\alpha, M_n}(1,2,3,4;5),
\eeqn{Psi12345-i}
where $M_d$ and $M_t$ are the projections of the deuteron and triton spins at large separations before the collision. The 3+2 partition is given by
\beq
\Phi^{(32)}_{dt, M_dM_t}(1,2,3;4,5) = \frac{1}{\sqrt{v_{dt}}}\frac{4\pi}{k_{dt}r_{dt}}
%\eol
%\times
\sum_{
{\scriptsize \begin{array}{ll} 
 {L'_{dt}S'_iL_{dt}S_iJ_i} \\ M_{dt} M_iM'_iM_{J_i} 
    \end{array}  } }i^{L_{dt}} 
\eol
%\sum_{L_{dt} M_{dt} S_i M_i J_i M_{J_i}}  
%(J_t M_t J_d M_d | S_i M_i) (L_{dt} M_{dt} S_i M_i | J_i M_{J_i})  
\eol
\times \,
%\sum_{S'_i M'_i}  
(J_t M_t J_d M_d | S_i M_i) \, [\psi_{t}(1,2,3 ) \otimes \psi_d(4,5) ]_{S'_i M'_i} %\sum_{ L'_{dt} M'_{dt} }
\,\,\,\,\,
\eol
\times \,
(L_{dt} M_{dt} S_i M_i | J_i M_{J_i})  
(L'_{dt} M'_{dt} S'_i M'_i | J_i M_{J_i})  \,\,\,\,\,\,\,\,\,\,\,\,\,\,
  \eol
\times \,
Y^*_{L_{dt}M_{dt}}(\hat{\ve{k}}_{dt})Y_{L'_{dt} M'_{dt}}(\hat{\ve{r}}_{dt})\chi^{J_i}_{L_{dt}S_{dt},L'_{dt}S'_{dt}} (k_{dt},r_{dt}), \,\,\,\,\,\,\,\,\,\,\,\,
\eeqn{chi32}
with $k_{dt} = \sqrt{2\mu_{dt} E_{dt}}/\hbar$ ($\mu_{dt}$ and $E_{dt}$ are the reduced mass and centre-or-mass energy of the $d$-$t$ system), $\ve{r}_{dt}$ representing the radius-vector connecting centres of mass of deuteron and triton, and  $\psi_{t}(1,2,3 ) $ and $\psi_d(4,5)$ being the triton and deuteron wave functions, respectively. The deuteron wave function will be represented below by the $s$-wave component only: $\psi_d(\ve{r}_{np}) = \varphi_d(r_{np}) Y_{00}(\hat{\ve{r}}_{np})$.

The 4+1 partition function is given by 
\beq
\Phi^{(41)}_{dt, M_dM_t}(1,2,3,4;5)=    \frac{1}{\sqrt{v_{i}}} \frac{4\pi}{k_{dt}r_{\alpha n}}
\sum \psi_{\alpha}(1,2,3,4) \, 
\eol
\times
%\sum_{L_{dt} M_{dt} S_{dt} M_{dt} J_i M_{J_i}}  
\chi^{(\rm spin)}_{\dem \sigma_n}  (5)(J_t M_t J_d M_d | S_{dt} M_{dt}) (L_{dt} M_{dt} S_{dt} M_{dt} | J_i M_{J_i})  \,\,
\eol
\times
%\sum_{ L_{\alpha n} M_{\alpha n} }
(L_i M_i \dem \sigma_n | J_i M_{J_i})  i^{L_i} Y^*_{L_{dt}M_{dt}}(\hat{\ve{k}}_{dt}) \,\,\,\,\,\,\,\,\,\,
  \eol
\times
Y_{L_i M_i}(\hat{\ve{r}}_{\alpha n})\chi^{J_i}_{L_{dt}S_{dt},L_{i}} (k_i,r_{\alpha n}), \,\,\,\,\,\,\,\,\,\,\,\,\,\,\,\,\,\,\,\,\,\,\,\,\,\,\,\,
\eeqn{chi41}
where $\ve{r}_{\alpha n}$ is the neutron coordinate-vector with respect to the centre of mass of $\alpha$ and the sum runs on all angular momenta and their projections, $\{ J_i M_{J_i} L_{dt} M_{dt} S_{dt} M_{dt} \sigma_n L_{\alpha n} M_{\alpha n} \} $. The  $k_{dt}$ and $k_i$ in the $d$-$t$ and $\alpha$-$n$ partitions are related by the energy conservation, 
\beq
\frac{\hbar^2 k_{dt}^2}{2\mu_{dt}}+Q = \frac{\hbar^2 k_i^2}{2\mu_{\alpha n}}
%= \frac{\hbar^2 k_f^2}{2\mu_{\alpha n}}+E_{\gamma}
\eeqn{ki}
where  the $Q$-value is 17.59  MeV. 
The representation   (\ref{chi41}) also includes   the $v_i^{-1/2}$ factor, where $v_i = \hbar k_i/\mu_{\alpha n}$ is the velocity in the $\alpha$-$n$ partition. 
 Because  the entrance channel wave function  does not have  incoming flux in this partition, the asymptotics of   $\chi^{J_i}_{L_{dt}S_{dt},L_i} (k_i,r)$ contains outgoing waves only:
\beq
\chi^{J_i}_{L_{dt}S_{dt},L_i} (k_i,r) \rightarrow - \frac{i}{2}  S^{J_i}_{L_{dt}S_{dt},L_i} O_{L_{i}}(k_ir),  \,\,\,\,\,\,
\eeqn{asymalphan}
where $O_l = G_l+iF_l$, with $F_l$ and $G_l$ being the regular and irregular   Coulomb functions, and  the $S$-matrix element $S^{J_i}_{L_{dt}S_{dt},L_{i}} $ describes transitions from the $d$-$t$ channel  to the $\alpha$-$n$ channel  with  quantum numbers $L_i J_i$ and $S_i = 1/2$. 

\subsection{ Direct and exchange amplitudes}

Because of the antisymmetrization requirement the matrix element that determines the transition between initial and final states has   direct and exchange parts:
\beq
\la \Psi_f  | {\cal M}^{M\lambda}_\mu |\Psi_i  \ra
=
M^{M\lambda;\, \rm dir}_{\mu}(k_{\gamma}) + M^{M\lambda;\, \rm ex}_{\mu}(k_{\gamma}), \,\,\,\,\,\,\,\,\,\,
\eeqn{}
where
\begin{widetext}
\beq
M^{M\lambda;\, \rm dir}_{\mu}(k_{\gamma}) = 
\la \Phi^{(41)}_{f}  (1,2,3,4;5) | {\cal M}^{M\lambda}_\mu(k_{\gamma}) |\Phi^{(41)}_{i}(1,2,3,4;5)\ra
+ 2\sqrt{2}  
\la \Phi^{(41)}_{f}(1,2,3,4;5) | {\cal M}^{M\lambda}_\mu(k_{\gamma}) |\Phi^{(32)}_{i}(1,2,3;4,5)\ra, \eol
\eeqn{ME1dir}
and
\beq
M^{M\lambda;\, \rm ex}_{\mu}(k_{\gamma}) = -4 \la \Phi^{(41)}_{f}(1,2,3,4;5) | {\cal M}^{M\lambda}_\mu(k_{\gamma}) |\Phi^{(41)}_{i} (1,2,3,5;4)\ra
+ 3\sqrt{2} 
\la \Phi^{(41)}_{f}(1,2,3,4;5) | {\cal M}^{M\lambda}_\mu(k_{\gamma}) |\Phi^{(32)}_{i}(3,4,5;1,2)\ra. \eol
\eeqn{ME1ex}
\end{widetext}
In the exchange part the nucleons 4 and 5 in the  4+1 component of the initial wave function are swapped while the final state wave function keeps 4 and 5 in the original order. In the 3+2 component of the entrance channel wave function the nucleons 1 and 2 are swapped with 4 and 5.

\subsection{M1 transition due to convection current}

The convection--current-induced part ${\cal M}^{M\lambda; \, \rm conv}_{\mu}$ of the M1 operator  does not contain any spin. Therefore, it does not cause any M1 transitions  from the strong $S=3/2$ component in the entrance channel to the $\alpha$-$n$ final state with the spin   $S=1/2$. This  term   connects the $d$-wave $S=1/2$ state of the 4+1  partition in the entrance channel with the $d$-wave final  $\alpha$-$n$ state. The motivation behind considering this contribution is that it can add to increase of the $d+t \rightarrow \alpha+n+\gamma$ cross section at low $\gamma$-energy contributed to bremsstrahlung, discussed in \cite{our}.
 
The formal expression for the ${\cal M}^{M\lambda; \, \rm conv}_{\mu}$  amplitude can be obtained  easily by relating 
the $\left[ \vec{\nabla}_j \phi_{\lambda \mu}(\ve{r}_j)\right] \cdot \vec{L}_j $ operator in the first term of (\ref{operatorM}) to the $\tfrac{1}{i}\left[ \vec{\nabla}_j \times \vec{L} \phi_{\lambda \mu}(\ve{r}_j)\right] \cdot \vec{\nabla}_j $ operator from  the convection-induced   $E\lambda$ transition amplitude  and then using developments from \cite{our}.
 Using (\ref{Lj}) and (\ref{philm})  one  obtains
\beq
\left[ \vec{\nabla}_j \phi_{\lambda \mu}(\ve{r}_j)\right] \cdot \vec{L}_j 
=
%
%   Uncomment to see derivation :
%
%\left[ \vec{\nabla}_j \phi_{\lambda \mu}(\ve{r}_j)\right] \cdot (-i \ve{r}_j \times \vec{\vec{\nabla}}_j )
%\eol
%=
%-i\left[ \vec{\nabla}_j \phi_{\lambda \mu}(\ve{r}_j)\right] \times  \ve{r}_j \cdot \vec{\vec{\nabla}}_j  
%=
%i\ve{r}_j \times \left[ \vec{\nabla}_j \phi_{\lambda \mu}(\ve{r}_j)\right]   \cdot \vec{\vec{\nabla}}_j  
%\eol
%=
-  \left[ \vec{L}_j \phi_{\lambda \mu}(\ve{r}_j)\right]   \cdot {\vec{\nabla}}_j  .
\eeqn{}
Since this operator should be sandwiched between the wave functions with a cluster structure, 
 $A = A_1 + A_2$, the operator ${\cal M}^{M\lambda; \rm conv}_{\mu}$ could be rewritten in a form similar to that from  \cite{Bay85}:
\beq
{\cal M}^{M\lambda; \rm conv}_{\mu} &=&
\frac{e\hbar}{m_p c} \frac{A}{A_1 A_2}\frac{(2\lambda+1)!!}{(\lambda+1) k_{\gamma}^{\lambda}}
\left[  Z_1 \tilde \chi_{\lambda \mu}\left( \frac{A_2}{A} k_{\gamma},\ve{r}\right) \right.
\eol
& &+\left.
(-)^{\lambda}Z_2 \tilde \chi_{\lambda \mu}\left( \frac{A_1}{A} k_{\gamma},\ve{r}\right)\right] \cdot \vec{\nabla}_{\ve{r}}.
\eeqn{MEcluster}
Here, $Z_i$ are the charges of the clusters $A_i$, $\ve{r} = \ve{R}_{\rm c.m.}(A_1) - \ve{R}_{\rm c.m.}(A_2)$ is the intercluster coordinate, connecting the centers of mass  positioned at $\ve{R}_{c.m.}(A_i)$, and  $\tilde \chi_{\lambda \mu}= -    \vec{L} \phi_{\lambda \mu}$.

The direct contribution to the convection part of the $M\lambda$ transition  amplitude is obtained by using Eq. (\ref{llnl}) from Appendix A and
making standard Racah algebra manipulations with summation over the angular momentum projections. For $\alpha$-$n$ bremsstrahlung (with $A_2=1$ and $Z_2=0$) it results in 
\beq
\la\Phi^{(41)}_{n\alpha, M_{n}}(1,2,3,4;5) \mid {\cal M}^{M\lambda; \, \rm conv}_{\mu}  \mid \Phi^{(41)}_{dt, M_d M_t}(1,2,3,4;5)\ra
\eol
=
\frac{e\hbar}{m_p c} \frac{(2\lambda+1)!!}{(\lambda+1) k_{\gamma}^{\lambda}}
\sum_{  L_{dt} S_i J_i   L_fJ_f }  {\cal R}^{(M\lambda);\, \rm conv,\, dir}_{L_{dt} S_i J_i L_f J_f}
%%\hat{L}'_i \hat{\lambda}\hat{J}_i  (L'_i 0 \lambda 0 | L'_f 0) 
%%W(\lambda L'_iJ_f \dem ;L'_fJ_i) 
% \sum_{L'_i  L'_f} \frac{(4\pi)^{3/2}}{k_ik_f}C^{(\lambda); \, \rm dir, \, conv}_{L'_iJ_iL'_fJ_f} I^{\rm conv}_{L_iL'_iJ_i,L_fL'_fJ_f;\lambda}(k_{\gamma})i^{-L'_f}i^{L'_i}
\,\,\,\,\,\,\,\,\,\,\,\,\,\,\,
\eol
\times
 \sum_{M_iM_{J_i}M_fM_{J_f} }
 Y_{L_fM_f}(\hat{\ve{k}}_f) 
 Y^*_{L_{dt}M_{dt}}(\hat{\ve{k}}_{dt})
 \,\,\,\,\,\,\,\,\,\,\,\,\,\,\,\,\,\,\,\,\,\,\, \,
\eol
\times \,
(L_f M_f \dem M_{n} | J_f M_{J_f}) (L_{dt} M_{dt} S_i M_i | J_i M_{J_i}) \,\,\,\,\,\,\,\,\,\,\,\,\,\,\,\,\,\,\,\,
 \eol
\times  \,( \lambda \mu J_iM_{J_i} | J_f M_{J_f})(J_t M_t J_d M_d | J_i M_{J_i}),  \,\,\,\,\,\,\,\,\,\,\,\,\,\,\,\,\,\,\,\,\,\,\,\,\,\,\,\,\,\,
\eeqn{M1convdir}
where
\beq
{\cal R}^{(M\lambda); \rm conv,dir}_{L_{dt} S_i J_i L_f J_f} =  \frac{Z_1 A}{A_1 A_2}  \frac{(-)^{\lambda}}{\sqrt{v_fv_i}}\frac{(4\pi)^{3/2}}{k_{dt}k_f}
%\hat{L}_i 
%
% Attention!  \hat_{L}_i will be replaced by \hat{L}_f because of change in definition of {cal D}
%
\hat{\lambda} \hat{J}_i
\sum_{  L_i L'_f  }  i^{L_i -L'_f}\hat{L}'_f
\eol
\times
%{\cal I}^{(\lambda); \, \rm conv, \, dir}_{L_{dt} S_iL'_iJ_i,L_fL'_fJ_f} (k_{\gamma}), \,\,\,\,
W(\lambda L_iJ_f\dem;L'_fJ_i)
\left(
{\cal D}^{(1)}_{L_i \lambda L'_f}
I^{(1)}_{\beta;\lambda}(k_{\gamma}) +{\cal D}^{(2)}_{L_i \lambda L'_f}
I^{(2)}_{\beta;\lambda}(k_{\gamma}) \right)
\eol
\eeqn{calR_convdir}
and contains the radial integrals
\beq
%I^{(1)}_{L_{dt} S_iL'_iJ_i,L_fL'_fJ_f;\lambda}(k_{\gamma})
I^{(1)}_{\beta;\lambda}(k_{\gamma})
=
\int_0^{\infty} dr  \, r \chi^{J_f}_{L_fL'_f} (k_f,r)   j_{\lambda}(\tfrac{1}{5}k_{\gamma}r) 
\eol
\times
\frac{\partial}{\partial r}  \frac{\chi^{J_i}_{L_{dt} S_iL'_i} (k_i,r)}{r} ,
\eeqn{I1}
and
\beq
%I^{(2)}_{L_{dt} S_iL'_iJ_i,L_fL'_fJ_f;\lambda}(k_{\gamma})
I^{(2)}_{\beta ;\lambda}(k_{\gamma})
=
\int_0^{\infty} \frac{dr}{r}  \chi^{J_f}_{L_fL'_f} (k_f,r)  j_{\lambda}(\tfrac{1}{5}k_{\gamma}r)   \chi^{J_i}_{L_{dt} S_iL'_i} (k_i,r) .
\eol
\eeqn{I2}
In this equations $\beta$ stands for the set of quantum numbers $\{L_{dt} S_iL'_iJ_i,L_fL'_fJ_f;\lambda\}$ and the coefficients ${\cal D}^{(1)}_{l_i \lambda l_f}$ and ${\cal D}^{(2)}_{l_i \lambda l_f}$ are given by Eq. (\ref{D1}) and (\ref{D2}) from the Appendix A, respectively.
To get formal expressions for the exchange amplitude, which contains contributions from relatively small $\alpha$-$n$ separations, the Siegert approximation to the amplitude (\ref{M1convdir}) could be used, in a similar fashion to what was  done in \cite{our}. However, as  shown below, the direct convection-induced M1 transitions are small enough to not worrying about the exchange effects.

\subsection{M1 transition due to magnetization current}

The magnetization current can connect $s$-wave states in  partitions 3+2 and 4+1  with different spin. The intrinsic  wave functions $\psi_\alpha$ and $\psi_t$ of $\alpha$-particle and triton in the corresponding two different partitions of $^5$He contain different numbers of nucleons. To deal with this situation, fractional parentage expansion is used here to separate one nucleon from $\alpha$-particle in the exit channel:
\beq
\psi_{\alpha} (1,2,3,4)&\approx& \, I_{tp}(\vec{r}_{pt})  \,[\psi_t(1,2,3)\otimes \chi_p(4)]_{00} 
\eol
&+& I_{nh}(\vec{r}_{nh}) [\psi_h(1,2,3)\otimes \chi_n(4)]_{00} 
\eeqn{psial}
which, according to standard practice,  includes  isospin Clebsch-Gordan coefficients into the radial parts of the overlap integrals $I_{tp}(\vec{r}_{pt}) = I_{tp}(r_{pt} ) Y_{00}(\hat{\vec{r}}_{pt})$ and $ I_{nh}(\vec{r}_{nh})=I_{nh}(r_{nh} ) Y_{00}(\hat{\vec{r}}_{nh})$. For  $d$-$t$ collisions, only the $t+p$ part of this expansion contributes to the M1 amplitude.

Using developments from the Appendix B and C one can show that
 the  ${\cal M}^{M1; \, \rm magn, \, dir}_{\mu}$ amplitude is given by Eq. (\ref{M1convdir}) with $\lambda=1$, $S_i=3/2$ and with ${\cal R}^{(M\lambda);\,  \rm conv,\, dir}_{L_{dt} S_i J_i L_f J_f}$ replaced by 
\beq {\cal R}^{(M\lambda);\, \rm magn,\, dir}_{L_{dt} S_i J_i L_f J_f}
=
-\frac{1}{4} \frac{k_{\gamma}}{\sqrt{v_{dt}v_f}} \frac{(4\pi)^{3/2}}{k_{f}k_{dt}} 
 (-)^{J_i+\lambda-J_f}
 \eol
 \times
 \sum_{L'_{dt}\lambda' L'_f S'_i } i^{L'_{dt}-L'_f}
 (-)^{L'_{dt}+\lambda'-L'_f} (-)^{\dem+S'_i} \sqrt{2} \hat{J}_i \hat{\lambda}'\hat{L}'_f    \hat{L}'_{dt}
\eol
\times\, 
c_{\lambda \lambda'}
{\cal I}_{L'_{dt}\lambda'L'_f} (k_{\gamma})
\left\{\begin{array}{lll} 
{L'_{dt}}&  {S'_i} & {J_i} \\
{\lambda'} & {1} & {\lambda} \\
{L'_{f}}&  {\dem} & {J_f} 
\end{array}\right\}, \,\,\,\,\,\,\,\,\,\,\,\,\,\,\,\,\,\,\,\,
\eeqn{RM1magn}
where
\beq
{\cal I}_{L'_{dt}\lambda'L'_f} (k_{\gamma})=\left( \tfrac{8}{5}\right)^3\int_0^{\infty}  d{r}_{\alpha n} r_{\alpha n}  \chi^{J_f}_{L_{f}L'_{f}} (k_{f},r_{\alpha n})\,\,\,\,\,\,\,\,\,\,\,\,\,\,\,
\eol
\times
\int_0^{\infty} d{r}_{dt} r_{dt}   F_{L'_{dt}\lambda'L'_f} (r_{dt},r_{\alpha n})   \chi^{J_i}_{L_{dt}S_iL'_{dt}S'_i} (k_{dt},r_{dt})\,\,\,\,\,\,\,\,\,\,\,\,\,\,\,
\eeqn{Icalmagn}
and $ F_{L'_{dt}\lambda'L'_f} (r_{dt},r_{\alpha n}) $ is given in the Appendix C. As for the exchange contribution to M1 magnetization amplitude, Appendix D shows that its leading term is zero.

\subsection{Differential cross section}

The differential cross section describing the
 $\gamma$-spectrum from $d$-$t$ collisions is obtained here following  Ref. \cite{Doh13}, which results in the expression  
 \beq
\frac{d\sigma}{dE_{\gamma} }&=& \frac{E_{\gamma}k^2_f}{\pi^2 \hbar^3 c}
\frac{1}{(2J_d+1)(2J_t+1)}\sum_{M_dM_t M_n}\frac{1}{4\pi}\int d\ve{\hat{k}}_{dt}  d\ve{\hat{k}}_f
\eol
&\times&
\sum_{\lambda\mu} \frac{1}{2\lambda+1}
\left|\alpha^{M\lambda}\la \Psi_{n\alpha, M_n } |
{\cal M}^{M\lambda}_{\mu}  | \Psi_{dt, M_d M_t}\ra\right|^2\,\,\,\,
\eeqn{dsigdEg}
with 
\beq
\alpha^{M\lambda} = -\frac{\sqrt{2\pi(\lambda+1)} i^{\lambda+1} k_{\gamma}^{\lambda}}{\sqrt{\lambda(2\lambda+1)}(2\lambda-1)!!}.
\eeqn{alpha}
%We can readily use these expressions because the normalization of (\ref{chi41}) and (\ref{chif}) are consistent with that adopted in \cite{Doh13}.
As usual, the unpolarized cross section contains summation over spin projection $M_n$ in the final state and averaging over the spin projections $M_d$ and $M_t$ of the colliding deuteron and triton  in the entrance state. The matrix element $\la \Psi_{n\alpha, M_n } |
{\cal M}^{M\lambda}_{\mu}  | \Psi_{dt, M_d M_t}\ra$ is given by the same partial expansion as in Eq.(\ref{M1convdir}) but with ${\cal R}^{(M\lambda); \, \rm conv, \, dir}_{L_{dt} S_i J_i L_fJ_f}$ replaced by
\beq
{\cal R}^{(M\lambda)}_{L_{dt} S_i J_i L_fJ_f} (k_{\gamma} )&=&{\cal R}^{(M\lambda); \, \rm conv, \, dir}_{L_{dt} S_i J_i L_fJ_f} (k_{\gamma} )
\eol
&+&2\sqrt{2}\,{\cal R}^{(M\lambda); \, \rm magn, \, dir}_{L_{dt} S_i J_i L_fJ_f} (k_{\gamma} ),
\eeqn{}
where the factor of $2\sqrt{2}$ arises due to the antisymmetrization, see Eq. (\ref{ME1dir}). After integrating over  directions of the final state momentum $\ve{k}_f$ and averaging over the incident momentum direction  $\hat{\ve{k}}_{dt}$, followed by summation over angular momentum projections, one obtains
\beq
\frac{d\sigma}{dE_{\gamma} } &=& \frac{e^2}{2\pi^2 } \frac{ k_{\gamma}k^2_f}{  m_p^2 c^2}
\sum_{L_{dt} S_i J_i L_fJ_f}
  \frac{2J_f+1}{(2J_d+1)(2J_t+1)}
\eol
&\times&
\sum_{\lambda}  \frac{ 1  }{ \lambda(\lambda+1)}      \left|{\cal R}^{(M\lambda)}_{L_{dt} S_i J_i L_fJ_f}(k_{\gamma})\right|^2.
     \eeqn{dsdEg}
%In our calculations we use $L_{dt}=0, S_{dt} = 3/2, L_f = 1$  and $J_i = 3/2$ so that the radial integral ${\cal R}_{q.n.}^{(\lambda)}(k_{\gamma})$ depends on $J_f$ only, which could be either 3/2 or 1/2.

\section{Numerical results}

Numerical calculations for M1 transitions have been performed here for convection-induced bremsstrahlung between two $d$-wave states of the $\alpha$-$n$ partition and for the direct transition from the $s$-wave $d$-$t$ continuum state to the $s$-wave $\alpha$-$n$ state caused by magnetization current. In both cases the final $\alpha$-$n$ wave function has been calculated from the potential model of \cite{Sat68} with Perey-correction for nonlocality of the range of 0.85 fm \cite{Fie66}, in the same way as it was done in \cite{our}. The introduction of the Perey factor is needed to simulate the five-body aspect of the $\alpha$-$n$ collision which would force the channel function to be a solution of a nonlocal equation.   As for the entrance-state wave functions, they are described below separately  for convection and magnetization M1 transitions.

Since all experimental data on $d+t \rightarrow \alpha+n+\gamma$ reaction are given as branching ratios to the cross section of the $\gamma$-less reaction $d+t \rightarrow \alpha+n$, results of all model calculations are also presented by this quantity, denoted as $\sigma_{dt\gamma}(E_{\gamma})/\sigma_{dt}$.

\subsection{Convection-induced bremsstrahlung}

 Following \cite{our}, to model the intermediate $d$-wave $\alpha$-$n$ channel function    the $\alpha$-$n$ potential from \cite{Sat68} has been renormalized in this wave to obtain exactly the same phase shifts as those given by ab-initio calculation in \cite{Nav12}. Then the $\alpha$-$n$ channel wave function was constructed by propagating the outgoing-wave-only asymptotic  solution  of the Schr\"odinger equation with the renormalized potential into the nuclear potential interior using the Numerov method.

Fig. 1 shows the $d$-wave bremsstrahlung spectrum calculated for $E_{dt}$ =  50 keV. Since no resonances are present in the   low-energy  $\alpha$-$n$  $d$-wave scattering, the spectrum has one broad feature only. It also exhibits a sharp increase towards zero energy $E_{\gamma}$, which  gives a divergent contribution to the integrated cross section. This is similar to the sharp increase of the  $d+t \rightarrow \alpha+n+\gamma$ cross sections in E1 transitions,  calculated in \cite{our} and attributed there to  unsuitability of the first-order perturbation theory at very small $E_{\gamma}$.   The branching ratio, $\sigma_{dt\gamma}(E_{\gamma})/\sigma_{dt} = 5.15 \times 10^{-8}$ integrated above the lowest point $E_{\gamma} $ = 0.85 MeV    is about three orders of magnitude smaller than the E1 value, which according to all measurement and previous theoretical calculations is around $5\times 10^{-5}$. Therefore, convection-induced transitions can be safely neglected. The same conclusion is drawn after several modifications of the intermediate channel $\alpha$-$n$ wave functions in the same fashion as described in \cite{our}. The model sensitivity was not sufficiently high to warrant further investigation of this M1 path.

\begin{figure}[t]
\includegraphics[scale=0.34]{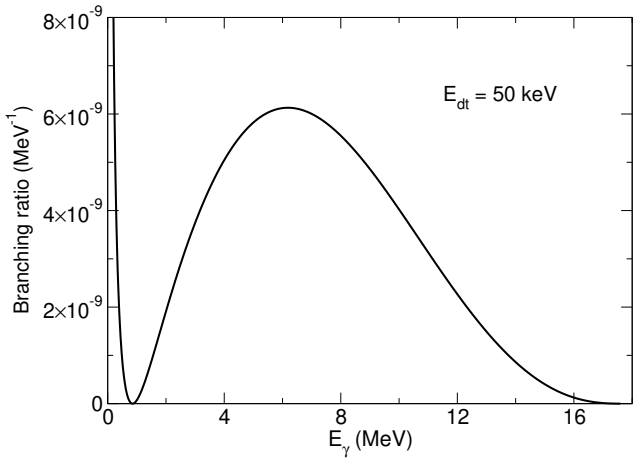}
\caption{The differential branching ratio  $\sigma_{dt\gamma}(E_{\gamma})/\sigma_{dt}$ for convection-current-induced M1 transition for the $d$-$t$ centre-of-mass energy $E_{dt}=50$ keV.}
\label{fig:M1conv}
\end{figure}

\subsection{Results for magnetization }

In a sharp contrast with convection-induced M1 transition, the magnetization path turned out to be very sensitive to the model choice for the $d$-$t$ scattering wave function. The first calculation, performed here, used the potential model from Dubovichenko et al \cite{Dub19}, originally designed  for a mirror scattering system, $d-^3$He. The $d-^3$He potential is represented in this model by one gaussian only, $V(r) = -V_0\,  e^{-0.1 r^2} $ with $V_0 = 34.5$ MeV. When applied to $d$-$t$ scattering, it gives the position of the 3/2$^+ $ resonance  at about 50 keV, which is close to the experimental value. With the   $d$-$t$  wave function at this energy the   maximum of the M1 spectrum is also located around 16.7 MeV, but is almost twice  the maximum of the E1 peak, while   the integrated branching ratio, 4.1$\times 10^{-5}$, is  comparable to that of  the E1 transition. This is   unrealistically large. 

As a next step, the optical model potential from Wu et al  \cite{Wu22} was used to generate the $d$-$t$ channel function. This potential is represented by a square well  with a range of 3.616 fm   both for the real and imaginary parts, with the corresponding real and imaginary well depths of    30 MeV and  49.64 keV, respectively. The corresponding  differential branching ratio $\sigma_{dt\gamma}(E_{\gamma})/\sigma_{dt}$, also calculated at 50 keV, is compared in Fig. \ref{fig:M1magn} to that obtained with the wave function from Dubovichenko et al. It is significantly smaller for  $E_{\gamma}> 7$  MeV, with the integrated  branching   ratio becoming 1.3$\times 10^{-6}$, which is   only  about 3$\%$ of the  E1 value. However,  in the area of the 17-MeV peak the M1 transition is almost half that of the E1 (see gray dashed line in Fig. \ref{fig:E1-M1}), which still seems to be very unrealistic.

\begin{figure}[t]
\includegraphics[scale=0.32]{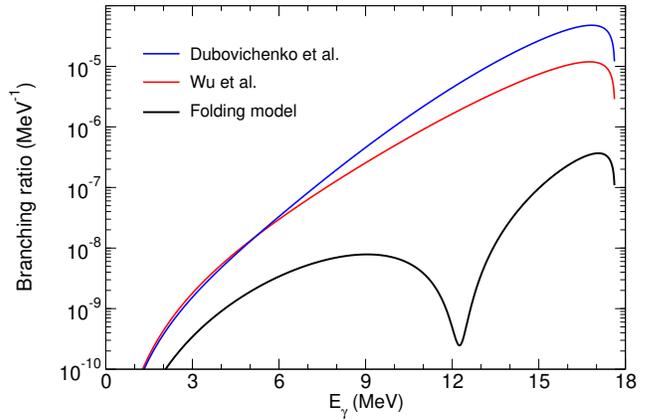}
\caption{The differential branching ratio  $\sigma_{dt\gamma}(E_{\gamma})/\sigma_{dt}$ for magnetization-induced M1 transition caused obtained with three models of the $d$-$t$ wave function, described in the main text, for the $d$-$t$ centre-of-mass energy $E_{dt}=50$ keV. }
\label{fig:M1magn}
\end{figure}

The M1 transition amplitude depends on how well the $\alpha$-$n$ and $d$-$t$ wave functions overlap in the integrand of Eq. (\ref{Icalmagn}). Because $ \chi^{J_i}_{L_{dt} S_iL'_f} (k_i,r_{dt})$ has a node, the ${\cal I}_{L'_{dt} \lambda' L'_f}(k_{\gamma})$ becomes very sensitive to cancellations between contributions from internal and external regions of the $d$-$t$ wave function.
The potential from Dubovichenko et al   has unrealistically fast (gaussian) decrease with increasing $d$-$t$ separations, while the potential from Wu et al  is just a step function representing a sudden disappearance of the $d$-$t$ interaction beyond a certain point,
%Both the Dubovichenko et al and Wu et al $d-t$ model potentials have unrealistically fast change in the   surface  area of the $d-t$ potential, 
which could have affected the strength of the M1 transition. The deuteron is a weakly bound object, so that one can expect that its interaction potential with other nuclei could be characterized by a large diffuseness. Therefore, a folding model was employed here to construct the $d$-$t$ potential at 50 keV. A simple deuteron wave function given by the Hulth\'en model 
\cite{Yam54} was chosen   to calculate  the deuteron density, which guarantees its correct asymptotic form. The triton density was presented by a single gaussian with the range of 1.37 fm chosen to provide the triton charge radius of 1.67 fm \cite{deV87}. The nucleon-nucleon interaction is modelled by the same Hulth\'en potential, which has a realistic rate of decrease. Then it was assumed that both the real and  imaginary  parts of the optical potential have the same shape. Their depths have been fitted to provide the $d$-$t$ phase shift of  16 degrees, close to the result from ab-initio calculations in \cite{Nav12},
%that reproduce within the experimental error bars the $d+t \rightarrow \alpha+n$ maximum,  
and the reaction cross section of 4.3 b, corresponding to the astrophysical $S$-factor of 27 MeV b. 
With this optical folding model the calculated $\sigma_{dt\gamma}(E_{\gamma})/\sigma_{dt}$  ratio  becomes 7.7$\times 10^{-5}$, which is about 1$\%$ of the E1 value integrated above $E_{\gamma}=10$ MeV. The differential branching ratios from all three $d$-$t$ interaction models are compared to each other  in Fig. \ref{fig:M1magn}. The minimum around 12 MeV in the folding model calculation is due to the almost complete cancellation between internal and external contributions from the $d$-$t$ wave function at this point. In Fig. \ref{fig:E1-M1} the M1 contribution calculated in the present optical folding model is compared to the E1 contribution from \cite{our}. It gives about 2$\%$ at the main peak but is negligible everywhere below 15 MeV. It starts to dominate the E1 transition above $E_{\gamma} \gtrsim 17.5$ MeV leading to additional counts in the $\gamma$-spectrum in this area, which potentially could be mistaken for background. 

\begin{figure}[t]
\includegraphics[scale=0.32]{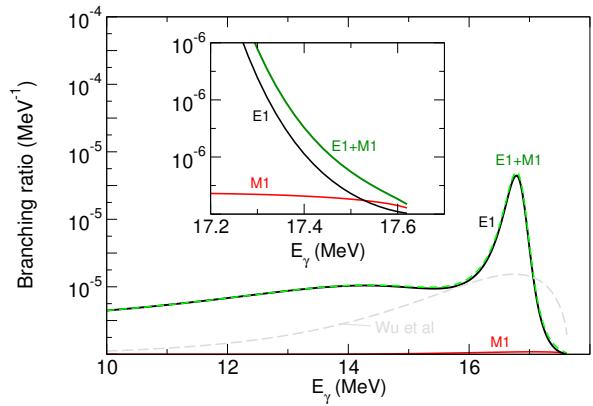}
\caption{The differential branching ratio  $\sigma_{dt\gamma}(E_{\gamma})/\sigma_{dt}$ for magnetization-induced M1 transition, calculated at $E_{dt}=50$ keV with optical folding model $d$-$t$ potential from this work,   in comparison to the E1 contribution from \cite{our}. The grey dashed line represents M1 calculation with optical model $d$-$t$ potential of Wu et al.}
\label{fig:E1-M1}
\end{figure}

To understand the difference between the three model calculations the wave functions from gaussian model of Dubovichenko et al, square well optical model of Wu et al and the present optical folding model are compared to each other in Fig. \ref{fig:wfs}. It is noticeable that the folding model wave function spreads out much further, with the node  at a significantly larger $d$-$t$ separation. This leads to a smaller absolute values of the wave function as compared to Dubovichenko et and and Wu et al, which is responsible for the small M1 contribution.

\begin{figure}[t]
\includegraphics[scale=0.32]{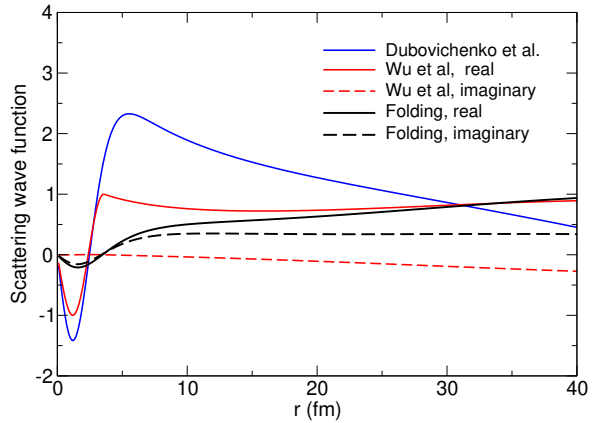}
\caption{ The $d$-$t$ scattering wave function at $E_{dt}=50$ keV calculated with gaussian potential from Dubovichenko et al, square well optical model from Wu et al and the optical folding model of the present paper. }
\label{fig:wfs}
\end{figure}

\section{Discussion and conclusions}

The present study of the M1 contribution to the $d+t \rightarrow \alpha+n+\gamma$ cross section shows that it is dominated by the direct transition from the $s$-wave $d$-$t$ partition of $^5$He to the $s$-wave $\alpha$-$n$ partition in the final channel.
The corresponding cross section has been found to be strongly dependent on the model choice for the $d$-$t$ scattering wave function. In particular, it depends on how fast the $d$-$t$ interaction potential decreases with the $d$-$t$ separation. With the choice of two potential models from the literature, one having a gaussian decrease and another one being just a square well, the M1 transition becomes unrealistically large due to the large values of $d$-$t$ channel function inside the $d$-$t$ interaction. The optical folding potential, constructed in this work, provides a more spatially extended $d$-$t$ interaction, which leads to a further spread    and the correspondingly smaller values of the $d$-$t$ wave function within the $d$-$t$ interaction region. This leads to a much smaller M1 cross section.

The strong model dependence of the M1 transition is reminiscent of the situation with theoretical calculations of $\beta$-delayed deuteron emission $^6$He $\rightarrow \alpha+d+e^-+\bar{\nu}_e$. Its measured branching ratio to the main $^6$Li ground state decay branch   is very small, of the order of 10$^{-6}$, however, potential model predictions are two orders of magnitude higher \cite{Bor93}. To bring theoretical predictions down, full three-body calculations of both the $^6$He $ = \alpha+n+n$ and $^6$Li $ = \alpha+n+p$ must be performed \cite{Tur06,Tur18},  keeping all the quantum numbers included. Such a situation is due to the  spatial parts of $^6$He and $\alpha+d$ wave functions being almost orthogonal, which could be achieved in full calculations only. 

The operator $\sum_i \sigma_it_{3i}$ that causes the Galow-Teller transition $^6$He $\rightarrow \alpha+d+e^-+\bar{\nu}_e$   is almost identical  to the M1 operator of this paper if the long-wave approximation is done.  Although all results for $d+t \rightarrow \alpha+n+\gamma$ shown here do not use the long-wave approximation, it has been checked numerically that this approximation does not change much the results. A similarity with the $\beta$-delayed deuteron emission of $^6$He is also in the three-body nature of the $d$-$t$ system, which could be considered as a strongly-bound triton plus neutron plus proton. The lessons learned from $\beta$-decay studies tell us that for  reliable calculations of the M1 transition one has to go beyond the two-body potential model description of the $d$-$t$ motion and include into consideration such details as deuteron breakup and polarization in the sub-Coulomb $d$-$t$ scattering. Also, more quantum numbers, such as the $d$-wave state admixture in the incoming deuteron, should be included in the calculations.

This work has been motivated by a possibility of having more counts near the endpoint of the $\gamma$-spectrum from $d$-$t$ collisions due to M1 transition to the final near-zero energy $s$-wave $\alpha$-$n$ state, which could be favored because of the absence of the centrifugal barrier. A realistic model of the $d$-$t$ interaction suggests that the strength of the spectrum in this area is about 2$\%$ only, which would not induce  much of a systematical error in the background subtraction. However, to confirm this conclusion,  as a minimum, three-body treatment of the $d$-$t$ collisions are required. In the long term, five-body calculations starting from realistic nucleon-nucleon interactions will be needed to clarify this matter.

\begin{acknowledgments}
       This work was part-funded by AWE, UK Ministry of Defence \textcopyright Crown owned copyright 2024/AWE. Remaining part of funding came from the United Kingdom Science and Technology Facilities Council (STFC) under Grant  No.   ST/V001108/1.
\end{acknowledgments}

\section{Appendix}

\subsection{Matrix element of $\vec{L} \phi_{\lambda \mu} \cdot \vec{\nabla}$}
To develop a formal expression  for M1 amplitude given by Eq.  (\ref{M1convdir}) one needs the following matrix element:
\beq
\la l_f m_f \mid \vec{L} \phi_{\lambda \mu} \cdot \vec{\nabla} \mid  l_i m_i \ra 
=
\la l_f m_f \mid \vec{L} \phi_{\lambda \mu} \cdot \vec{n} \mid  l_i m_i \ra \frac{\partial}{\partial r}
\eol
+ \, 
\la l_f m_f \mid \vec{L} \phi_{\lambda \mu} \cdot \vec{\nabla}_{\Omega} \mid  l_i m_i \ra \frac{1}{r}. \,\,\,\,\,\,\,\,\,\,\,\,\,\,\,\,\,\,\,\,
\eeqn{Lpn}
Using
\beq
L_{\nu} \phi_{\lambda \mu} = \sqrt{\lambda(\lambda+1)}\, j_{\lambda}(kr) \sum_{\mu'} (\lambda \mu 1 \nu | \lambda \mu') Y_{\lambda \mu'} (\hat{\ve{r}}) \,\,\,\,\,\,\,\,\,\,\,\,\,\,\,\,
\eeqn{}
and
\beq
\la l_f m_f \mid\vec{\nabla}_{\Omega} \mid  l_i m_i \ra
=
(l_i m_i \lambda\mu | l_f m_f) \hat{l}^{-1}_f  {\cal C}_{l_i l
_f},
\eeqn{}
\beq
 {\cal C}_{l_i l_f } = -\left( l_i \sqrt{l_i+1} \delta_{l_f,l_i+1}   + (l_i+1) \sqrt{l_i} \delta_{l_f,l_i-1} \right),
 \,\,\,\,\,\,\,\,\,\,\,
\eeqn{calC}
one obtains
\beq
\la l_f m_f \mid \vec{L} \phi_{\lambda \mu} \cdot \vec{\nabla} \mid  l_i m_i \ra 
=
\hat{\lambda}\, (l_i m_i \lambda\mu | l_f m_f) %\hat{l}^{-1}_f 
\eol
\times
%
% Attention! removing this factor will change eq. (18) where\hat_{L}_i will be replaced by \hat{L}_f
%
\frac{%\hat{l}_i 
1}{\sqrt{4\pi} %\hat{l}_f
}j_{\lambda} (k r)
\left( {\cal D}^{(1)}_{l_i \lambda l_f} \frac{\partial}{\partial r} + {\cal D}^{(2)}_{l_i \lambda l_f}\frac{1}{r}\right) ,
\eeqn{llnl}
where
\beq
{\cal D}^{(1)}_{l_i \lambda l_f}
%=
%-\sqrt{\lambda(\lambda+1) } \, \sum_L (l_i 0 1 0 | L 0) 
%\eol
%\times
%(L 0 \lambda 0 | l_f 0) W(l_i 1 l_f \lambda; L\lambda)
%\eol
%=-\sqrt{\lambda(\lambda+1) } \, \sum_L ( L 0 1 0 |l_i   0) \hat{L}\hat{l}_i^{-1} (-)^1
%\eol
%\times
%(  l_f 0 \lambda 0 |L 0) (-)^{\lambda} \hat{l}_f \hat{L}^{-1} W(l_i 1 l_f \lambda; L\lambda)
%\eol
=
(-)^{\lambda} \sqrt{\lambda(\lambda+1) }
%\,  \hat{l}_f  \hat{l}_i^{-1}  
%
% Attention! removing this factor will change eq. (18) where\hat_{L}_i will be replaced by \hat{L}_f
%
\sum_L ( L 0 1 0 |l_i   0) 
\eol
\times
(  l_f 0 \lambda 0 |L 0)  W(l_i 1 l_f \lambda; L\lambda)
\eeqn{D1}
and 
\beq
{\cal D}^{(2)}_{l_i \lambda l_f}
=
(-)^{\lambda+1} \sqrt{\lambda(\lambda+1) (2\lambda+1)}  \,\, %\hat{l}_f  \hat{l}^{-1}_i 
%
% Attention! removing this factor will change eq. (18) where\hat_{L}_i will be replaced by \hat{L}_f
%
\eol
\times
\sum_L  {\cal C}_{l_i L }(l_f 0 \lambda 0 | L 0) W(l_i 1 l_f \lambda; L\lambda).
\eeqn{D2}

\subsection{Spin and isospin matrix elements}

The contribution from nucleons inside the triton is characterized by $e^{i\vec{k}_{\gamma}\vec{r}_j}$. The coordinate of each nucleon is $\vec{r}_j = \vec{r}'_j - \vec{r}_t$, where  $\vec{r}'_j$ is its position with respect to the triton centre of mass and $\vec{r}_t= -\tfrac{2}{5}\vec{r}_{dt}$ is the position of the latter with respect to the $^5$He centre of mass.  Because triton is spatially confined  one can assume  that $e^{i\vec{k}_{\gamma}\vec{r}_j}\approx 1$.
In this approximation the M1 magnetization-induced amplitude does not contain spatial variables and, therefore, one has to consider matrix elements in the spin space only. There are three such matrix elements, the first of which is
\beq
\sum_{i=1,2,3} \la \psi_t (1,2,3) \mid S_{\mu, i} g_{s_i} \mid \psi_t(1,2,3) \ra  \,\,\,\,\,\,\,\,\,\,\,\,\,\,\,\,\,\,\,\,\,\,\,\,\,\,\,\,\,\,\,\,\,\,\,\,
\eol
=\frac{1}{2}(g_n+g_p) \la  \psi_t  \mid S_{\mu} \mid \psi_t\ra \,\,\,\,\,\,\,\,\,\,\,\,\,\,
\eol
+ \,(g_n-g_p) \sum_{i=1,2,3} \la \psi_t (1,2,3) \mid S_{\mu, i} t_{3i} \mid \psi_t(1,2,3) \ra \,\,\,\,\,\,\,\,\,\,\,\,
\eeqn{spinme}
To evaluate the second term it is assumed here that the triton wave functions is dominated by the fully antisymmetric in spin-isospin space component $ \left| [3]  S=\dem M_S T=\dem M_T \right. \ra$ described by the Young's tableau $[f]=[3]$ with spin and isospin equal to 1/2 and the spin and isospin projections of $M_S$ and $M_T$, respectively. Using fractional parentage expansion to extract the $m$-th nucleon from this spin-isospin component:
\beq
 \mid [3]  S=\dem M_S T=\dem M_T \ra = 
 \eol
 \sum_{[f_2]S_2 T_2 M_2 M_{T_2}} \la [3]  \dem \dem \mid [f_2] S_2   T_2  \ra \mid [f_2] S_2 M_2 T_2M_{T_2} \ra
 \eol
\times (S_2 M_2 \dem \sigma | \dem M_S)  (T_2 M_{T_2} \dem \tau | \dem M_{T}) \chi^{\rm spin}_{\dem \sigma} (m)\chi^{\rm isopin}_{\dem \tau} (m),
 \eol
 \eeqn{}
where $\la [3]  \dem \dem \mid [f_2] S_2   T_2  \ra $ is the fractional parentage expansion coefficient and $\mid [f_2] S_2 M_2 T_2M_{T_2} \ra$ is the two-body spin-isospin function, and then performing all summations over spin and isospin projections, one obtains
\beq
\la [3]  \dem M'_S \dem M'_T \mid \sum_{i=1,2,3}S_{\mu, i} t_{3i}  \mid 
[3]  \dem M_S \dem M_T \ra
\eol 
=
3\la [3]  \dem M'_S \dem M'_T \mid  S_{\mu, m} t_{3m}  \mid 
[3] \dem M_S \dem M_T \ra
\eol
\eol
= M_T \delta_{M_T M'_T } \la \psi_t | S_\mu | \psi_t \ra.
\,\,\,\,\,\,\,\,\,\,\,\,\,\,\,\,\,\,\,\,\,\,\,\,\,\,\,
 \eeqn{}
Therefore, for the isospin projection $M_T=1/2$, associated with triton,
\beq
\sum_{i=1,2,3} \la \psi_t (1,2,3) \mid S_{\mu, i} g_{s_i} \mid \psi_t(1,2,3) \ra = 
\eol
%(\dem(g_n+g_p) + (g_n-g_p)M_T) \la \psi_t | S_\mu | \psi_t\ra 
%= \left(g_n \delta_{M_T,\dem}+  g_p\delta_{M_T,-\dem}\right) \la \psi_t | S_\mu | \psi_t\ra .
= g_n
\la \psi_t | S_\mu | \psi_t\ra  =\frac{1}{ \sqrt{2} } g_n (\dem M'_t  1 \mu | \dem M_t)  \la \dem || \ve{S} || \dem \ra .
\,\,\,\,\,\,\,\,\,\,\,\,
\eeqn{}
Then, supplementing the triton wave function with  spin functions of theremaining neuton and proton, one has
\beq
\la [[\psi_t \otimes \chi^{\rm spin}_{\dem} (p)]_0 \otimes \chi^{\rm spin}_{\dem} (n)]_{\dem \sigma_n}  |  
\sum_{i=1,2,3}  {S}_{\mu,i} g_{si} 
\eol
 \times \, |   \chi^{\rm spin}_{\dem}
[ \psi_t \otimes \chi^{\rm spin}_{1} (p,n)]_{S_i M_{S_i}} \ra
%=\frac{3}{2\sqrt{2}}\hat{S}_i (1 \mu S_i M_{S_i}| \dem \sigma_n) W(1 \dem \dem 1;\dem S_i)
%\eol
%=
%\frac{3}{2\sqrt{2}}\sqrt{4}(1 \mu S_i M_{S_i}| \dem \sigma_n) (1/3) 
\eol
= g_n\frac{1}{\sqrt{2}} (1 \mu S_i M_{S_i}| \dem \sigma_n).
\,\,\,\,\,\,\,\,\,\,\,\,\,\,\,\,\,\,\,\,\,\,\,\,
\eeqn{i123}

For $i=4$, which is a proton in deuteron,   the spin part is simpler,
\beq
\la [[\psi_t \otimes \chi^{\rm spin}_{\dem} (p)]_0 \otimes \chi^{\rm spin}_{\dem} (n)]_{\dem \sigma_n}  |  {S}_{\mu}(p) g_{sp}
\eol
\times
| [ \psi_t  \otimes \chi^{\rm spin}_{1} (p,n)]_{S_i M_{S_i}} \ra
%\eol
%=
%\sum_{S  } (-)^{1/2-S_i-S}\hat{S}^2 \frac{1}{\sqrt{2}}\sqrt{\frac{3}{2}}
%W(\dem \dem \dem \dem; 0S ) W(\dem \dem S_i \dem; 1 S) 
%\sqrt{3}
%(1 \mu S_i M_{S_i}| \dem \sigma_n) W(1 \dem \dem S; \dem S_i) \sqrt{2} \hat{S}_i
%\eol
%=
% (-)^{1/2-3/2-1} 3 \frac{1}{\sqrt{2}}\sqrt{\frac{3}{2}}
%\frac{1}{2}\frac{1}{3}
%\sqrt{3}
%(1 \mu S_i M_{S_i}| \dem \sigma_n) \frac{1}{3} \sqrt{2} \sqrt{4}
%\eol
=
g_p \sqrt{\frac{1}{2}}(1 \mu S_i M_{S_i}| \dem \sigma_n),  
\,\,\,\,\,\,\,\,\,\,\,\,
\eeqn{i4}
 which is exactly the same as in the case of $i=1,2,3$ except for the neutron gyromagnetic factor $g_n$ being replaced by the the proton gyromagnetic factor $g_p$.

For $i=5$, which is a neutron in deuteron,   the spin  matrix element is
\beq
\la [[\psi_t \otimes \chi^{\rm spin}_{\dem} (p)]_0 \otimes \chi^{\rm spin}_{\dem} (n)]_{\dem \sigma_n}  |  {S}_{\mu}(n) g_{sn}
\eol
\times
| [ \psi_t  \otimes \chi^{\rm spin}_{1} (p,n)]_{S_i M_{S_i}} \ra
\eol
=
W(\dem \dem S_i;01) \frac{1}{2} \la \dem || \vec{S} || \dem \ra (\dem M_{S_i} 1 \mu | \dem \sigma_{n_f}),
\eeqn{}
which is non-zero only for $S_i=1/2$. Given that the M1 transitions considered in this paper come from the resonance component with spin $S_i=3/2$, the contribution from nucleon 5 (which is the neutron in deuteron) is set to zero. 

Finally, the matrix elements (\ref{i123}) and (\ref{i4})  should be further multiplied by the isospin part of the M1 matrix element, which is
\beq
\la  \chi^{\rm isospin}_{\dem \dem}(1,2,3)   \chi^{\rm isospin}_{\dem -\dem}(4)    \chi^{\rm isospin}_{\dem \dem} (n)  
\,\,\,\,\,\,\,\,\,\,\,\,\,\,\,\,\,\,\,\,\,\,\,\,\,\,\,\,\,\,\,\,\,
\eol
\times \,
|    \chi^{\rm isospin}_{\dem \dem}(1,2,3)  \chi^{\rm isospin}_{00} (p,n)  \ra
=
(\dem -\dem \dem \dem | 00) \eol
= -\frac{1}{\sqrt{2}}.
\,\,\,\,\,\,\,\,\,\,\,\,\,\,\,\,\,\,\,\,  
\eeqn{}

\subsection{Spatial matrix element for M1 magnetization amplitude }

The calculate the M1 amplitude the integration should be performed either over $d\vec{r}_{\alpha n} d\vec{r}_{pt}$ or over $d\vec{r}_{np} d\vec{r}_{dt}$.
This task is simplified if
% $d\vec{r}_{\alpha n} d\vec{r}_{pt}$ to 
integration is performed over $d\vec{r}_{\alpha n} d\vec{r}_{dt}$ - the variables of the continuum state wave functions. Because $\vec{r}_{tp} = \tfrac{8}{5} \vec{r}_{dt} - \tfrac{4}{5} \vec{r}_{\alpha n} $ and $\vec{r}_{np} = \tfrac{8}{5} \vec{r}_{\alpha n} - \tfrac{6}{5} \vec{r}_{dt} $, one has $d\vec{r}_{\alpha n} d\vec{r}_{pt}=\left(\tfrac{8}{5}\right)^3\vec{r}_{\alpha n} d\vec{r}_{dt}$. Then   the product of the radial parts of the overlap $I_{pt} $ and the deuteron wave function $\psi_d$ could be represented as
\beq
I_{pt}({r}_{pt})\varphi_d({r}_{np})
 &=& 4\pi \sum_{lm} {\cal F}_{l}(r_{dt},r_{\alpha n}) Y^*_{lm}(\hat{\ve{r}}_{dt})Y_{lm}(\hat{\ve{r}}_{\alpha n }),
\eol 
{\cal F}_{l}(r_{dt},r_{\alpha n}) &=& \frac{1}{2}
 \int_{-1}^1 d\mu \, P_{\lambda}(\mu) \varphi_d \left(|\tfrac{6}{5}\ve{r}_{dt} - \tfrac{8}{5}\ve{r}_{\alpha n} |\right)
 \eol
&\times&
 I_{pt}(|\tfrac{8}{5} \ve{r}_{dt} - \tfrac{4}{5} \ve{r}_{\alpha n}|).
\eeqn{expansion}
 Using the results for both spin and isospin matrix elements derived in the subsection above, one can obtain that  the M1 amplitude contains the matrix element over the angular variables
\begin{widetext}
\beq
F= -i \sum_{lm} {\cal F}_{l} (r_{dt},r_{\alpha n}) \la  Y_{L'_fM'_f}(\hat{\ve{r}}_{\alpha n}) Y_{lm}(\hat{\ve{r}}_{dt}) \mid g_n\left[\vec{\nabla}_t \times \vec{L}_t \phi_{\lambda \mu}(r_t)\right]_{\nu}+g_p \left[\vec{\nabla}_p \times \vec{L}_p \phi_{\lambda \mu}(r_p)\right]_{\nu} \mid  Y_{L'_{dt}M'_{dt}}(\hat{\ve{r}}_{dt})  Y_{lm}(\hat{\ve{r}}_{\alpha n}) \ra.
\eol
\eeqn{FF}
\end{widetext}
 The first term  is further developed using
\beq
 \la   Y_{lm}(\hat{\ve{r}}_{dt}) \mid \left[\vec{\nabla}_t \times \vec{L}_t \phi_{\lambda \mu}(r_t)\right]_{\nu}  \mid  Y_{L'_{dt}M'_{dt}}(\hat{\ve{r}}_{dt}) 
\ra
\eol
=k  
\sum_{\lambda' \mu'}(-)^{\lambda'} c_{\lambda\lambda'} (\lambda \mu 1 \nu |\lambda' \mu') j_{\lambda'}(\tfrac{2}{5} k_{\gamma} r_{dt}
)\eol
\times\,
(L'_{dt} 0 \lambda' 0 | L'_f 0) \frac{\hat{L}'_{dt} \hat{\lambda}'}{\sqrt{4\pi} \hat{L}'_f}
\,\,\,\,\,\,\,\,\,\,\,\,\,\,\,\,\,\,
\eeqn{}
 with
  \beq
  c_{\lambda \lambda'} &=& -(\lambda+1) \sqrt{\frac{\lambda}{2\lambda-1}},  \,\,\,\,\,\,\,\,  \lambda' = \lambda-1 \eol
 & &  0  \,\,\,\,\,\,\,\,\,\,\,\,\,\,\,\, \,\,\,\,\,\,\,\,\,\,\,\,\,\,\,\,\,\,\,\, \,\,\,\,\,\,\,\,\,\,\,\,\,\,\,\,\,\,\,\,\, \lambda'=\lambda \eol
  & & \lambda \sqrt{\frac{\lambda+1}{2\lambda+3}},  \,\,\,\,\,\,\,\, \,\,\,\,\,\,\,\,\,\,\,\,\, \,\,\,\,\,\,\,\,\lambda' = \lambda+1. 
   \eeqn{}
To develop the second term of (\ref{FF}}), one has to use a general expansion  for $\ve{r} = \alpha_1 \ve{r}_1 + \alpha_2 \ve{r}_2$
\beq
-i \left[\vec{\nabla} \times \vec{L} \phi_{\lambda \mu}(r)\right]_{\nu}  = 
\sqrt{4\pi} k \sum_{\lambda' \mu' \lambda_1 \lambda_2 \mu_1 \mu_2}i^{-\lambda_1+\lambda_2-\lambda' } 
%\hat{\lambda}_1 \hat{\lambda}' \hat{\lambda}^{-1}_2 
c_{\lambda \lambda'}
\eol
\times
(\lambda_1 0 \lambda_2 0 | \lambda' 0) j_{\lambda_1}(\alpha_1 k r_1) j_{\lambda_2}(\alpha_2 k r_2) 
\eol
\times
 (\lambda_1 \mu_1 \lambda' \mu' | \lambda_2 \mu_2) (\lambda \mu 1 \nu | \lambda' \mu )Y^*_{\lambda_1 \mu_1}(\hat{\ve{r}}_1)Y_{\lambda_2 \mu_2}(\hat{\ve{r}}_2)
 \eol
\eeqn{}
In the case of $\ve{r}_p = \tfrac{6}{5} \ve{r}_{dt} -\tfrac{4}{5} \ve{r}_{\alpha n}$ the final result for $F$ becomes
\beq
 F&=&
  k \frac{\hat{L}'_{dt} }{\sqrt{4\pi}} 
 \sum_{\lambda' \mu'}
 c_{\lambda \lambda'} F_{L'_{dt}\lambda'L'_f} (r_{dt},r_{\alpha n})
\eol
&\times&
(\lambda \mu 1 \nu | \lambda' \mu' )  (\lambda' \mu' L'_{dt} M'_{dt} | L'_fM'_f),
\eeqn{}
where
\beq
F_{L'_{dt}\lambda'L'_f} (r_{dt},r_{\alpha n})=
g_p 
{\cal G}_{L'_{dt}\lambda'L'_f} (r_{dt},r_{\alpha n})\,\,\,\,\,\,\,\,
\eol
+ \,
g_n   (L'_{dt} 0 \lambda' 0 | L'_f 0)j_{\lambda'}(\tfrac{2}{5}kr_{dt}) 
{\cal F}_{L'_f} (r_{dt},r_{\alpha n})
\eeqn{}
with
\beq
{\cal G}_{L'_{dt}\lambda'L'_f} (r_{dt},r_{\alpha n})
=
\sum_{\lambda_1 \lambda_2 l} {\cal C}_{\lambda_1 \lambda_2 l}{\cal F}_{l} (r_{dt},r_{\alpha n}) 
\eol
\times
j_{\lambda_1}(\tfrac{6}{5} k r_{dt}) j_{\lambda_2}(\tfrac{4}{5} k r_{\alpha n}) 
\eeqn{}
and
\beq
{\cal C}_{\lambda_1 \lambda_2 l}
=
\sum_{ \lambda_1 \lambda_2  }i^{\lambda_1+\lambda_2-\lambda' } \hat{\lambda}^2_1  \hat{\lambda}^2_2 
(\lambda_1 0 L'_{dt} 0 | l 0)  
\eol
\times
(\lambda_2 0 L'_{f} 0 |l  0) 
 (\lambda_1 0 \lambda_2 0 | \lambda' 0) W(\lambda'\lambda_1L'_fl; \lambda_2 L'_{dt}). 
 \eeqn{}
Then  standard algebra manipulations lead to (\ref{RM1magn}) and (\ref{Icalmagn}).

\subsection{Exchange contribution to magnetization}

The simplest way to calculate    the exchange contribution
\beq
\la \Phi^{(41)}_{\alpha n,M_n}(1,2,3,4;5)\mid {\cal M}^{M\lambda}_{\mu} \mid\Phi^{(32)}_{dt,M_d M_t}(3,4,5;1,2)\ra \nonumber
\eeqn{}
to the M1 amplitude is to use  the long-wave approximation, justified by smallvalues of the $k_{\gamma} r$ contributing to the process. In this approximation, one has to deals with matrix elements of the operator $ \sum_i g_{si} S_{\mu,i} = (g_n+g_p) \sum_i  S_{\mu,i}+ (g_n-g_p) \sum_i  S_{\mu,i}t_{3i}$. This could be achieved by making fractional parentage decomposition of the $\alpha$-particle wave function in the exit channel into the product of two deuteron wave functions keeping only the terms corresponding to the deuteron ground state wave function,
\beq
\psi_{\alpha}(1,2,3,4) \approx \left[\psi_d(1,2)\times \psi_d(3,4) \right]_{00} I_{dd}(\vec{r}_{dd}),
\eeqn{}
with the isospin Clebsch-Gordon coefficient embedded in the overlap function $I_{dd}(\vec{r}_{dd})$.  The triton wave function in the initial state is also expanded as
\begin{widetext}
\beq
\psi_{t}(3,4,5) &=& I_{nd}(\vec{r}_{nd}) \varphi_{d}(\vec{r}_{34})
[\chi^{\rm spin}_{1} (3,4)
 \chi^{\rm spin}_{\dem} (5) ]_{J_t M_t}
\eeqn{}
Then   considering  only the  $s$-wave components in the deuteron wave functions  and overlaps one has to calculate the matrix element
\beq
\la [\chi^{\rm spin}_{1} (1,2) \times  \chi^{\rm spin}_{1}  (3,4)]_{00} \,
\chi^{\rm spin}_{\sigma_{n_f}} (5)  X^{\rm isospin} 
%\chi^{\rm isospin}_{00} (1,2 ) \chi^{\rm isospin}_{00} (3,4)\chi^{\rm isospin}_{\dem \dem} (5)  
\mid 
\sum_{i=1}^5 S_{\mu,i} g_{s_i} 
\mid [   [\chi^{\rm spin}_{1} (3,4)\times \chi^{\rm spin}_{\dem} (5) ]_{J_t }\times 
\chi^{\rm spin}_{1}(1,2)]_{S_i M_{S_i}}
X^{\rm isospin} 
%\chi^{\rm isospin}_{0} (3,4)\chi^{\rm isospin}_{\dem \dem } (5) \chi^{\rm spin}_{0}(1,2)
\ra,\,\,\,\,\,
\eeqn{spin-isospin}
where $X^{\rm isospin} $is the isospin part of the five-body system. It is easy to show that
\beq 
\sum_{i=1,2} \la \chi^{\rm spin}_{1M'_d}(1,2)\chi^{\rm isospin}_{00}(1,2)
\mid S_{\mu,i}t_{3i} \mid \chi^{\rm spin}_{1 M_d}(1,2)\chi^{\rm isospin}_{00}(1,2)\ra=0 
\eeqn{}
and
\beq
\sum_{i=1,2} \la \chi^{\rm spin}_{1M'_d}(1,2)\chi^{\rm isospin}_{00}(1,2)
\mid S_{\mu,i}\mid \chi^{\rm spin}_{1 M_d}(1,2)\chi^{\rm isospin}_{00}(1,2)\ra
=
 -\sqrt{2} (1 \mu 1 M_d | 1 M'_d).  
\eeqn{}
The same is true for nucleons 3 and 4. Also, 
\beq
\la \chi^{\rm spin}_{\dem \sigma'_5} (5)\chi^{\rm isospin}_{\dem,\dem} (5) \mid S_\mu g_{s_5} \mid
\chi^{\rm spin}_{\dem \sigma_5} (5)\chi^{\rm isospin}_{\dem,\dem} (5)\ra = g_n (\dem \sigma_5 1 \mu | \dem \sigma'_5) \frac{1}{\sqrt{2}}\la \dem || S || \dem \ra
\eeqn{}
Then, after further algebraic manipulations, one obtains that the spin-isospin latrix element (\ref{spin-isospin}) is equal to
\beq
\frac{1}{2}(g_n+g_p)(\dem M_{S_i} 1 \mu | \dem \sigma_{n_f})
\sum_S  (-)^{S+S_i}  \hat{S}_i \hat{S} W(11\dem S ; 1 S_i) \left[ (-)^{J_t+1-S_i} \delta_{S J_t} + (-)^{3/2-S}\hat{S} \hat{J}_t  W(1\dem  S_i 1 ; S J_t) \right]
\eol
+
(-)^{J_t+1-S_i}g_n
  W(11S_i \dem; 0 J_t)   \hat{J}_t (\dem M_{S_i} 1 \mu | \dem \sigma_{n_f})
  \,\,\,\,\,\,\,\,\,\,\,\,\,\,\,\,\,\,\,\,\,\,\,\,\,\,\,\,\,\,
\eeqn{}
For $S_i = 3/2$ and $J_t = 1/2$ it is equal to zero.

\end{widetext}

\bibliographystyle{apsrev4-1}

\end{document}